\documentclass[preprint2]{aastex6}
\pdfoutput=1
\usepackage{times}
\usepackage{wasysym} 
\usepackage{hyperref} 
\usepackage{natbib} 
\usepackage{multirow}
\usepackage{graphicx}

\let\oldtabular\tabular
\renewcommand{\tabular}{\scriptsize\oldtabular}

\usepackage[normalem]{ulem}

\begin{document}

\title{Transiting Planets with LSST IV: Detecting Planets around White Dwarfs}
\author{Michael B.\ Lund\altaffilmark{1},
Joshua A.\ Pepper\altaffilmark{2},
Avi Shporer\altaffilmark{3},
Keivan G.\ Stassun\altaffilmark{1,4}}
\altaffiltext{1}{Department of Physics and Astronomy, Vanderbilt University, Nashville, TN 37235, USA; \\ \url{michael.b.lund@vanderbilt.edu}}
\altaffiltext{2}{Department of Physics, Lehigh University, Bethlehem, PA 18015, USA}
\altaffiltext{3}{Department of Physics, and Kavli Institute for Astrophysics and Space Research, Massachusetts Institute of Technology, Cambridge, MA 02139, USA}
\altaffiltext{4}{Department of Physics, Fisk University, Nashville, TN 37208, USA}


\begin{abstract}
Previous work has demonstrated that the Large Synoptic Survey Telescope (LSST) has the capability to detect transiting planets around main sequence stars in relatively short ($<$ 20 days) periods and using standard algorithms for transit detection and period recoverability. In this paper, we demonstrate how an algorithm proposed in \citet{Tingley2011} can be used for detecting transiting planets around white dwarfs with LSST. This application offsets the very short transit duration with the large change in magnitude caused by a transit of a white dwarf so that only a few points in transit are needed to detect periodicity and constrain the period. Our initial simulations find that approximately 1 in 5 close-in transiting planets would be detectable around white dwarf hosts; at an occurrence rate of 0.25 earth-sized planets per white dwarf, this is over 500 planets that can be detected. We also note that the current low cadence of LSST in the Galactic Plane has a significant impact on the expected yield.
\end{abstract}

\section{Introduction}
The search for exoplanets has, to large extent, been driven by the specific goal of finding planets as similar to the Earth as possible. Thus, missions such as \emph{Kepler} have focused their observations predominantly on Main Sequence stars, particularly with an emphasis on G dwarfs \citep{Batalha2010}. In particular, the parameter of interest from such surveys has been eta-Earth, the fraction of Sun-like stars that have Earth-like planets \citep{Catanzarite2011}. However, this has meant that other sorts of stars have not been targeted for planet searches to the same degree as Sun-like stars. Planets around red dwarfs have been viewed as attractive search candidates, particularly as the transit depth of an Earth-like planet is much larger around a smaller star, and the habitable zones occur much closer to the star, and consequentially, are easier to observe. Some of this work has been carried out by using the \emph{Kepler} observations of red dwarfs in the \emph{Kepler} field \citep{Dressing2015}, but there have also been dedicated searches of planets around red dwarfs, such as MEarth \citep{Nutzman2008}. Toward the other end of the main sequence, some transit surveys are discovering planets around hotter main sequence stars, such as KELT-9b around a 10,170 K star \citep{Gaudi2017} and Kepler-13b around a 7,650 K star \citep{Shporer2011, Shporer2014, Mazeh2012}. There have also been a small number of planets that have been discovered around pulsars, but not as part of a large-scale survey \citep{Wolszczan1992}.

Planets around white dwarfs, in contrast, have been primarily discussed only indirectly, with time-series photometric surveys being of much smaller scale than searches of main sequence stars. As close-in planets are expected to not survive the AGB stage of stellar evolution \citep{Villaver2007}, any planets in close-in orbits would likely require either new planet formation to occur close to the white dwarf or or mechanisms to move planets closer, and several methods for this to occur have been proposed \citep{Debes2002, Livio2005, Faedi2011, Veras2013}. White dwarfs are known to have disks of material around them \citep{Xu2012}, and polluted white dwarfs show signs of having accreted material from bodies that had undergone differentiation and are dominated by core-like or mantle-like material \citep{Harrison2018}. Additionally, simulations have shown that a tidally disrupted planet, such as a super-Earth, can result in a second-generation Earth-sized planet as it may be able to reform with a sizeable portion of the disk mass \citep{VanLieshout2018}. If close-in planets exist, then these systems may be the most favorable in searching for biomarkers \citep{Loeb2013a}, and some planets may at the correct temperatures for water (if present) to exist as a liquid on the surface, potentially for Gyrs \citep{Agol2011}.

There have already been some attempts or proposals to search for planetary bodies in orbit around white dwarfs. The broad requirements for an ideal search for planets around white dwarfs was discussed in \citet{Agol2011}, including a discussion of strategies for searching for transiting planets and also noting that if the occurence rate around white dwarfs is low, then a survey that observes many white dwarfs at once, such as LSST, will be need. \citet{Faedi2011} searched approximately 200 white dwarfs in the WASP (Wide-Angle Search for Planets) database for transit or eclipse events, and no evidence was found for transiting or eclipsing planetary companions. There was also a proposal to use the K2 mission to search habitable planets around white dwarf, by observing around 100 white dwarfs per field \citep{Kilic2013a}. Most notably, this yielded a white dwarf hosting at least one (and likely several) disintegrating planetesimals in a very short orbit around a 16,000 K white dwarf, WD 1145+017 \citep{Vanderburg2015}. Using the $\sim$1000 white dwarfs that were observed by \emph{Kepler}, \citet{VanSluijs2017} provided constraints on the number of white dwarfs with habitable zone Earth-sized planets at $<$28\%, and close-in Jupiters at a rate of less than 1.5\%. They also note that to further constrain these properties, a much larger sample of white dwarfs will be needed to suitably constrain exoplanet frequency around white dwarfs. While there has also been some discussion of Gaia's potential to astrometrically detect massive exoplanets, these would be planets that are in wide orbits, and very unlikely to transit \citep{Silvotti2014a}.

\begin{figure*}[!htb]
  \begin{center}
      \includegraphics[width=0.6\textwidth]{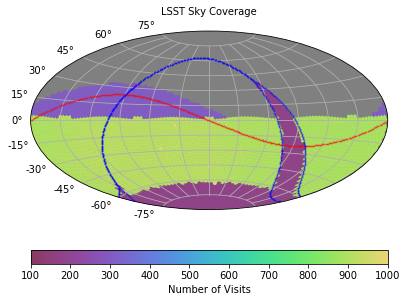}
  \end{center}
  \caption{The approximate number of observations per field under the current LSST observing strategy. The wide-fast-deep survey accounts for most of the sky, with $\sim$1000 observations. A select number of deep-drilling fields receive many more observations, and exceed the upper limit on the plot's color bar. Other regions including the Galactic plane receive 200-300 observations. This image is generated using the Metric Analysis Framework (MAF) that has been released for characterization of LSST \citep{Jones2014a}. (Available at \url{https://www.lsst.org/scientists/simulations/maf})}
  \label{fig:LSSTcounts}
\end{figure*}

The Large Synoptic Survey Telescope (LSST) is an 8.4-meter telescope currently under construction in Cerro Panchon, Chile \citep{Ivezic2008}. LSST will observe more than half the sky over the course of ten years, with most areas being observed $\sim$1000 times as part of the wide-fast-deep survey, but some areas (referred to hereafter as deep-drilling fields) will be observed on order of 10,000 times. Additionally, the Galactic plane, the South Celestial Pole, and the Northern portion of the ecliptic will all be observed only around 200-300 times. We show a representative sky map in Figure~\ref{fig:LSSTcounts} of the number of observations over ten years for each LSST field. These represent the current observing strategy for LSST; however this may change as a result of input from the LSST community \citep{LSSTScienceCollaboration2017}.
Each LSST observation, termed a visit, is made up of two 30-second exposures, and will take place in one of the six LSST bands, \emph{ugrizy}. Previous papers have demonstrated that LSST can contribute to our understanding of exoplanets by discovering planets in short periods around G, K and M main sequence stars using commonly-used transit search algorithms \citep{Lund2014, Jacklin2015, Jacklin2017}.

In this paper, we demonstrate that LSST can detect not just planets around main sequence stars, but also transiting planets around white dwarfs, if they exist in sufficient numbers. We note that we here focus on planets smaller than their host stars, and so we simulate these using the same framework as transiting planets and refer to them as transiting planets, but these could also be thought of as eclipsing or occulting planets, particularly in cases where the planet radius is equal to or larger than the stellar radius. We initially begin with a toy model of LSST in Section~\ref{sec:approx} to provide an approximate idea of how the likelihood of detecting a transiting planet around a white dwarf will depend on planet period. We then carry out a simulation of LSST light curves to develop a more robust yield. We outline our method for creating simulated light curves in Sections~\ref{sub:CatSim} and \ref{sub:lc}, and also demonstrate the usefulness of an algorithm for high signal to noise transit events in Section~\ref{sub:Tingley}. We show in Section~\ref{results} that with our assumptions, LSST would be able to detect 1 out of every 5 transiting planets, and if each white dwarf has a single close-in planet, then the brightest 3 million white dwarfs can yield over 2000 transiting exoplanets. At a lower rate of 0.25 planets per white dwarf, this still represents over 500 transiting exoplanets.

\section{Methods}
\subsection{Basic Approach}\label{sec:approx}
Traditional ground-based detection of transiting planets has centered on cases where the planet depth is on the order of 1\% or lower, which is comparable to the precision. This has meant that methods like box-least squares fitting, which check a range of periods to search for transit-like events, are needed to determine when a transit event has occurred. In contrast, a planet transiting a white dwarf represents a very significant dimming that is much larger than photometric precision. In this case, even a single point in transit represents a significantly statistical variability, and can identify a white dwarf as a potential host to a transiting planet.

\begin{figure}[!htb]
  \begin{center}
      \includegraphics[width=0.4\textwidth]{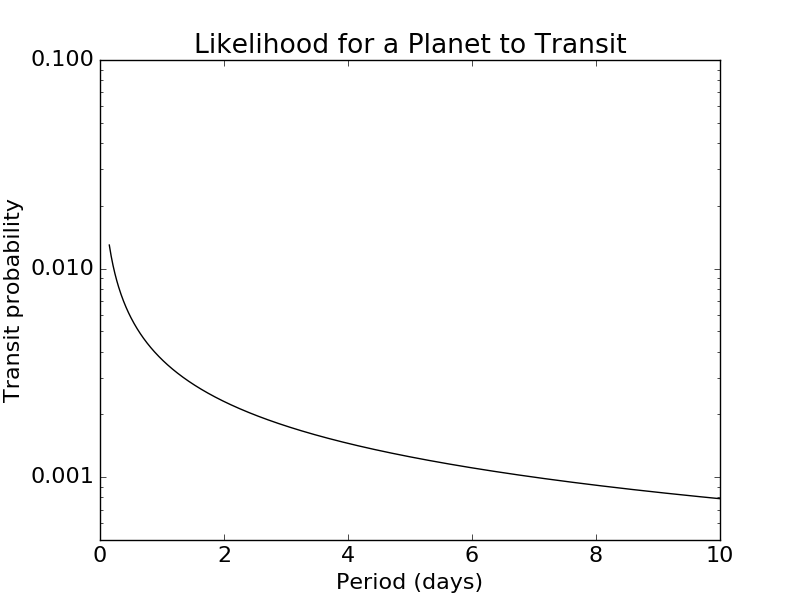}
  \end{center}
  \caption{The geometric probability of a planet being on a transiting orbit around a white dwarf as a function of period. The planet radius is neglected in this calculation.}
  \label{fig:analytic1}
\end{figure}

We first look at the transit probability for a planet around a white dwarf as a function of period, using a canonical mass and radius for a white dwarf of 0.6 solar masses and 0.013 solar radii, respectively. We examine the transit probabilities for planets orbiting close to the star, with orbital periods of 0.15 to 10 days and presuming circular orbits and equatorial transits. In Figure~\ref{fig:analytic1} we show the transit probability as a function of period, ignoring planet radius in this calculation. We note that in a case where the planet radius is close to the stellar radius, a significant portion of transiting planets will not be accounted for when we ignore the planet radius. However, the planets that we miss here are those with impact parameters greater than 1, and these planets will have grazing transits that will be shorter and shallower and so simulating these with a simple boxcar transit would overstate our results. We leave addressing transits at high impact parameters to future work. For the selected period range, the transit probability ranges from about 1\% for the closest planets, dropping to 0.1\% for the farthest planets.

\begin{figure}[!htb]
  \begin{center}
      \includegraphics[width=0.4\textwidth]{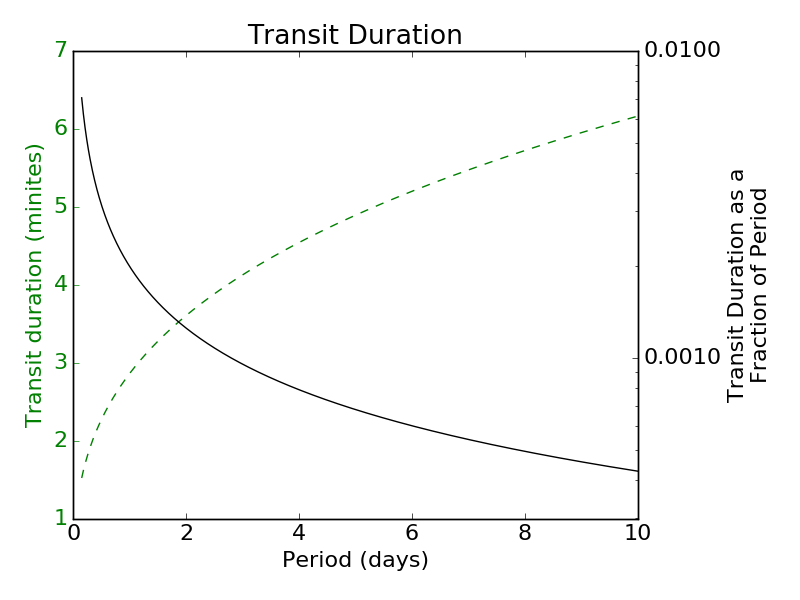}
  \end{center}
  \caption{Duration of the transit of an earth-sized planet as a function of period. Absolute duration in minutes is shown by the green dashed line, and duration as a fraction of the total period is the solid black line.}
  \label{fig:analytic2}
\end{figure}

In Figure~\ref{fig:analytic2}, we look at the transit duration, again only based on the stellar radius. The fraction of time that the planet is transiting drops from 0.7\% for the shortest-period planets, to 0.04\% for the longest-period planets. This is much less than the fractional durations of Hot Jupiter transits in this period range, which are on the order of several percent. The transit duration is even more different from known exoplanets, with transits that range from 1.5 to 6 minutes. Observing these short-duration transits would require that exposure times be kept short, but this also means that a full transit can be observed in under ten minutes in all cases of a transiting planet in a circular orbit with less than a 10 day period.

\subsection{LSST Toy Model}\label{sec:ToyModel}

\begin{figure*}[!htb]
  \begin{center}
      \includegraphics[width=0.4\textwidth]{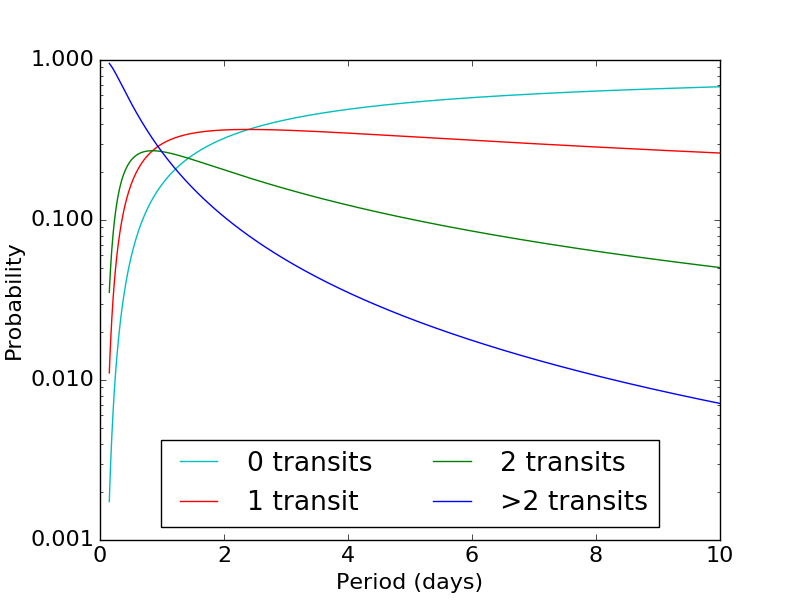}
      \includegraphics[width=0.4\textwidth]{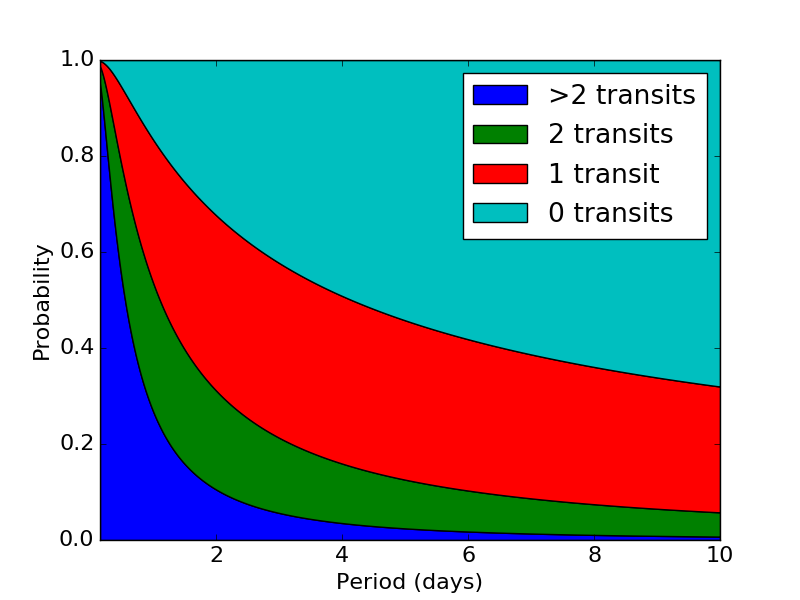}
  \end{center}
  \caption{The chance of a transiting planet having 0, 1, 2, or more than 2 points in transit when observed 900 times is shown in both figures. More than 2 points in transit are required for detection. On the left, we show the individual probabilities of each category, and on the right we show this as a stacked plot to better highlight the distribution of the number of transits as a function of period. In both cases, a planet never observed in transit is shown in cyan, those with 1 transit are in red, those with 2 transits are in green, and those with more than 2 transits are in blue.}
  \label{fig:analytic3}
\end{figure*}

We apply a model of LSST's observing schedule to the functions outlined in the previous section. Basing this only on the primary LSST observing mode (the wide-fast-deep survey), we look at the chance that LSST will observe white dwarf planets in transit. For most ground-based observations of planets transiting main sequence stars, the transit depth is roughly on order of the photometric noise or lower, but planetary transits of white dwarfs in LSST are much larger than the photometric noise. We show the probability of a planet having 0, 1, 2, and 3 or more LSST observations in transit in Figure~\ref{fig:analytic3}. Of particular interest here is that at short periods of less than about 1 day, a planet is more likely than not to have at least 3 points observed in transit over the nominal LSST mission, with the shortest periods having almost all transiting planets being observed at least 3 times in transit. At ten days, this number is still at around 1\%.

\begin{figure*}[!htb]
  \begin{center}
      \includegraphics[width=0.4\textwidth]{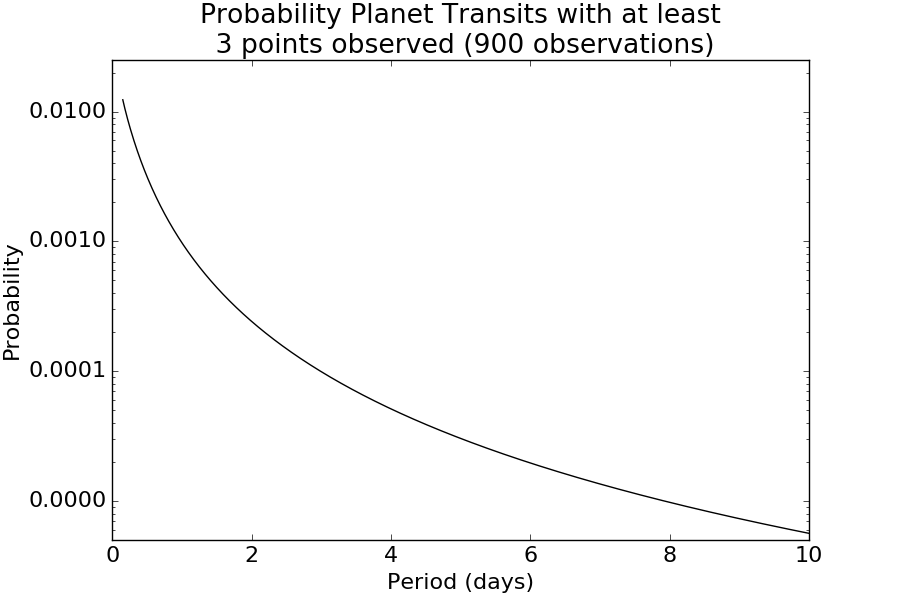}
      \includegraphics[width=0.4\textwidth]{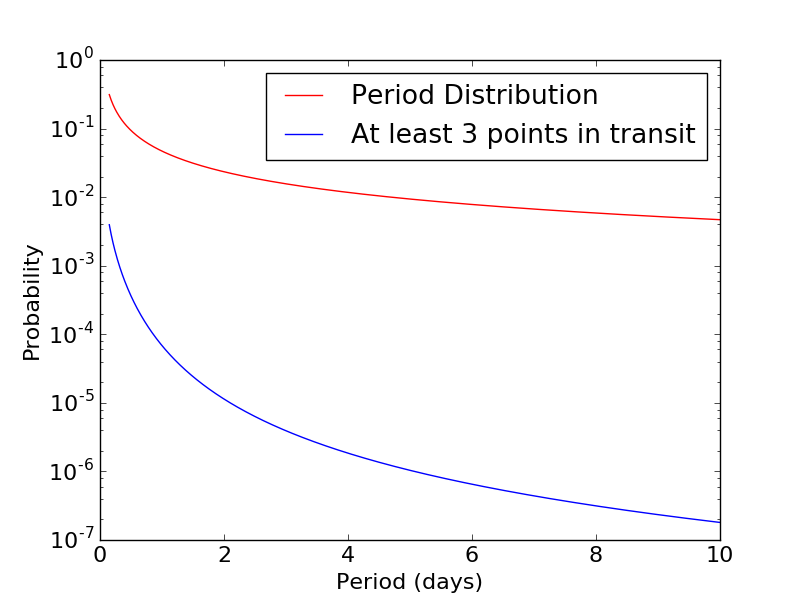}
  \end{center}
  \caption{On the left, the likelihood of a planet both transiting, and then being observed at least 3 times in 900 observations. On the right, a log-uniform period distribution is shown in red and this multiplied by the likelihood of a planet transiting and being observed at least three times is shown in blue.}
  \label{fig:analytic4}
\end{figure*}

We then convolve the transit likelihood with the chance that a transiting planet will be observed at least 3 times in transit in a set of 900 independent observations, which corresponds to the approximate number of observations in LSST's standard wide-fast-deep survey, and the results are shown on the left in Figure~\ref{fig:analytic4}. At short periods where transiting planets will be observed with high probability, this function is dominated primarily by the likelihood of a planet being on a transiting orbit, and is greater than 1\%. At long periods, this drops to about 0.001\% as both the probability of a transiting orbit and an in-transit observation become small.

To characterize, to first order, how many potentially recoverable planets there could be, we assume that each white dwarf hosts one planet, with possible periods represented by a log-uniform distribution ranging from 0.15 to 10 days. We then multiply this period distribution by the chance that a planet will transit as a function of period, and that a transiting planet would be observed at least three times during transit, to determine the likelihood that a white dwarf will have a planet that will be sufficiently observed to have its period constrained. In the right panel of Figure~\ref{fig:analytic4}, we show the probability distribution of initial planet periods in red, and in blue the probability for a planet that will be observed at least three times during transit. To determine the yield, we multiply the probability of a white dwarf having a detectable transiting planet in our period range by the number of white dwarfs that will be observed in LSST, which we put at 1.8 million, as described in Section~\ref{sub:CatSim}. Overall, our rough expectation is that there will be about 7,900 white dwarf observed by LSST with transiting planets, and of them, 4,500 will be observed at least 3 times in transit.
\vfill
\subsection{Simulated White Dwarf Population}\label{sub:CatSim}
To access a population of white dwarfs for our simulations, we use the LSST Catalog Simulator (CatSim\footnote{https://www.lsst.org/scientists/simulations/catsim}), which provides a complete simulation of all objects that can be observed by LSST \citet{Connolly2010, Connolly2014}. For stars within the galaxy, the distribution is drawn from the {\it Galfast} model of \citet{Juric2008}, which generated stars as members of the halo and thin and thick disk populations. The White dwarf spectral energy distributions are fit to LSST apparent magnitudes by way of \citet{Bergeron1995}, with Galactic reddening applied using the 3d Galactic model from \citet{Amores2005}. In our work, we then queried this already created catalog to obtain all white dwarfs in the simulation.

\begin{figure*}[!htb]
  \begin{center}
      \includegraphics[width=0.4\textwidth]{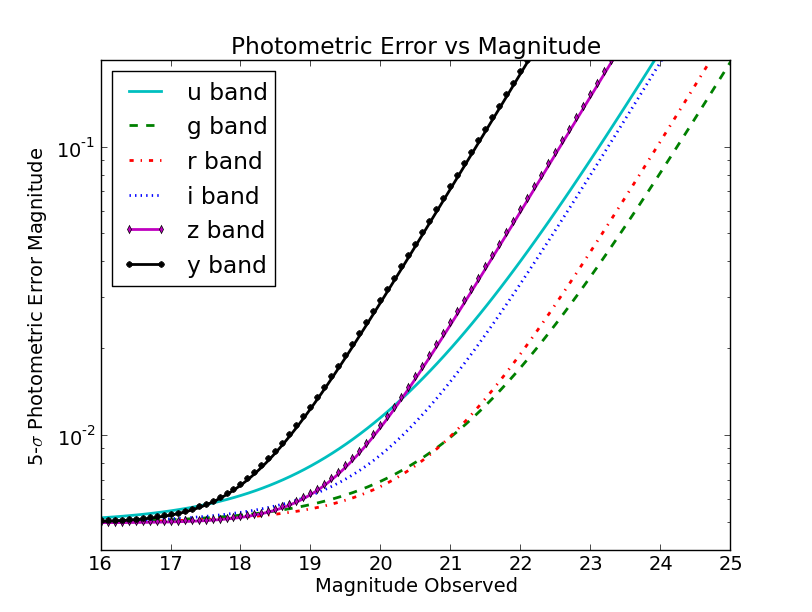}
      \includegraphics[width=0.4\textwidth]{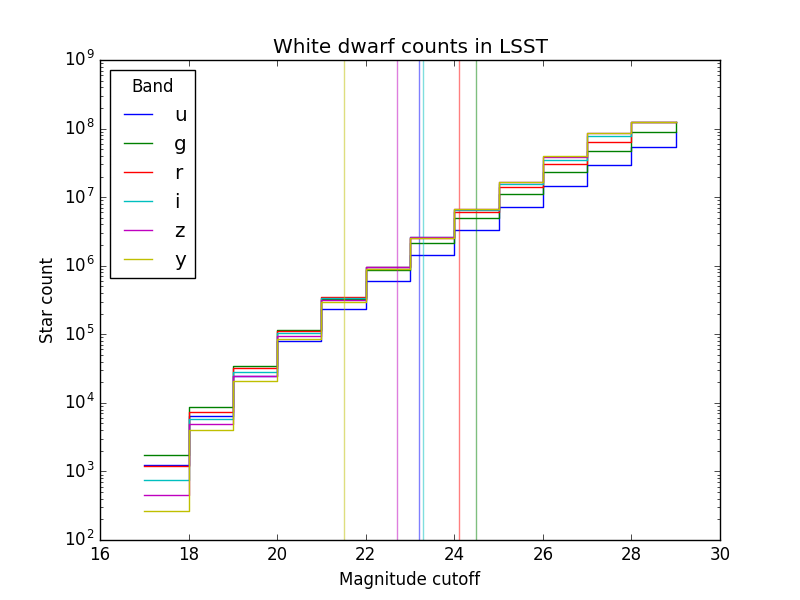}
  \end{center}
  \caption{The noise model for LSST on the left \citep{Ivezic2008}. On the right, the number of stars brighter than a given magnitude, in one magnitude bins. The vertical lines represent the magnitude for the noise model to have a 5-$\sigma$ photometric error of 0.1 mag.}
  \label{fig:catalog}
\end{figure*}

The current LSST CatSim database hosted by the University of Washington contains over 124 million white dwarfs, however we choose to focus only on a subset of the brightest white dwarfs. We consider only white dwarfs with an apparent magnitude of 23 or brighter in at least one of the LSST bands (with the exception of the \emph{y} band, where the magnitude cutoff is 22). We include the noise model for all six LSST bands in Figure~\ref{fig:catalog}. After we apply our cuts, we are left with 2667822 white dwarfs in the LSST footprint for our simulations. Of these, 1814953 are observed at at least the wide-fast-deep cadence, and 1170761 are observed at low cadence, and are primarily in the Galactic plane.
The LSST CatSim does not include the physical characteristics of the white dwarfs being simulated, only the observable parameters, and so we set the white dwarf masses and radii to 0.6 M$_{\odot}$ and 0.013 R$_{\odot}$ (1.4 R$_{\rm Earth}$), respectively.

\vfill
\subsection{Light Curve Simulation}\label{sub:lc}
The light curve simulations are carried out in a manner largely similar to that described in \citet{Lund2014}. We populate all white dwarfs in our sample with 1 Earth-radius planet each at a period drawn from a log-uniform distribution between 0.15 and 10 days. These planets are then assigned a random inclination to calculate if the planet would transit, assuming circular orbits and equatorial transits.

To obtain a time series for observations, we rely on the LSST Operations Simulator (OpSim) results that have been developed by the LSST team (we are specifically using OpSim v2.3.2, run 3.61 results). In order to quickly simulate the light curves, we ignore overlaps between LSST fields by only using the observations associated with the LSST field that a given star is closest to the center of. Therefore, our results serve as a lower limit on the transit yield by. While there are newer simulated LSST cadences than OpSim v2.3.2, run 3.61, it shares its broad characteristics with more recent OpSim results.

We then use the apparent magnitudes from CatSim to provide baseline magnitudes. For those stars hosting transiting planets, we add a simple boxcar transit with depth, period, and duration determined by the system parameters.
All light curves then have noise added to them in accordance with the LSST noise model \citep{Ivezic2008}.

\subsection{Periodicity Searching}\label{sub:Tingley}
Transits of planets around main sequence stars are generally no deeper than about 1\% in flux, and as this is often comparable in scale to the photometric precision of ground-based observations, it is often not possible to determine that a single observation represents a point in transit. However, even an Earth-sized planet transiting a white dwarf will cause a very significant dimming event and so transit detections will have very high signal to noise. For this reason, we use an algorithm proposed by \citet{Tingley2011} that is designed specifically for finding periodic transit signals in sparsely-sampled, long baseline data sets by focusing on high S/N outliers.

For a single light curve, we identify all observations that exceed 15 median average deviations (MAD) below the median and consider these outlying data points to be candidate observations of planetary transits. We then take the times of all candidate transit observations for that given light curve, and consider every possible pairing of two times. For each pair of times, we can define a set of ranges in period space that could contain the true period, as any two points in transit can be described as being separated by an integer multiple of the real period plus or minus the event duration. For a light curve with $n$ points in transit, we then have $n(n-1)$ sets of period ranges. As we are asserting that all outlier points must be points in transit, and we are assuming that there is only a single object transiting at a fixed period, we are then left with a solution that is the intersection of $n(n-1)$ sets of period ranges. A key point here is that the width of each period range is determined by the maximum duration of the event, and so an event with a very small duration will have very narrow ranges in period space.

As there are no known Earth-sized planets transiting white dwarfs, to provide an initial demonstration that this method works on real data, we look at a particular eclipsing binary system observed by the Kilodegree Extremely Little Telescope (KELT, \citet{Pepper2007}) that shares some of the key characteristics that this algorithm requires. 
Since the fractional event duration is an order of magnitude larger for this EB, the period ranges in our solution will also be much larger, and this will make our solution easier to display (a white dwarf case is presented in Section~\ref{results}, and the possible period ranges are much smaller). The eclipsing binary system we use only has a prominent primary eclipse, dimming about 0.15 magnitudes. This primary eclipse occurs every 32 days, but only lasts about 0.4 days. Much as the white dwarf case, this EB has a constant brightness for most of the time it is observed with an eclipse that is very deep compared to the noise in the light curve. Additionally, the eclipse represents only a small portion of the total period, in this case spending about 3\% of its time in eclipse as opposed to about 0.1\% of the time that a white dwarf spends in transit. 

\begin{figure*}[!htb]
  \begin{center}
      \includegraphics[width=0.4\textwidth]{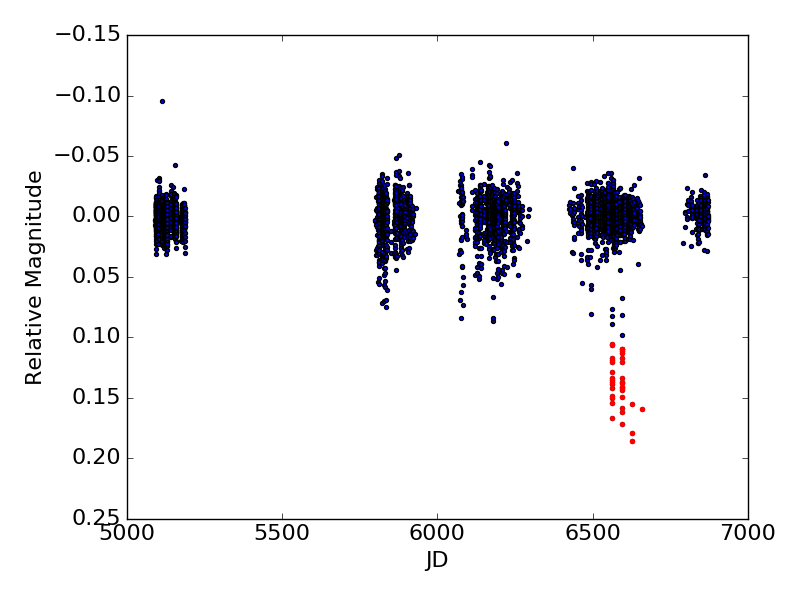}
      \includegraphics[width=0.4\textwidth]{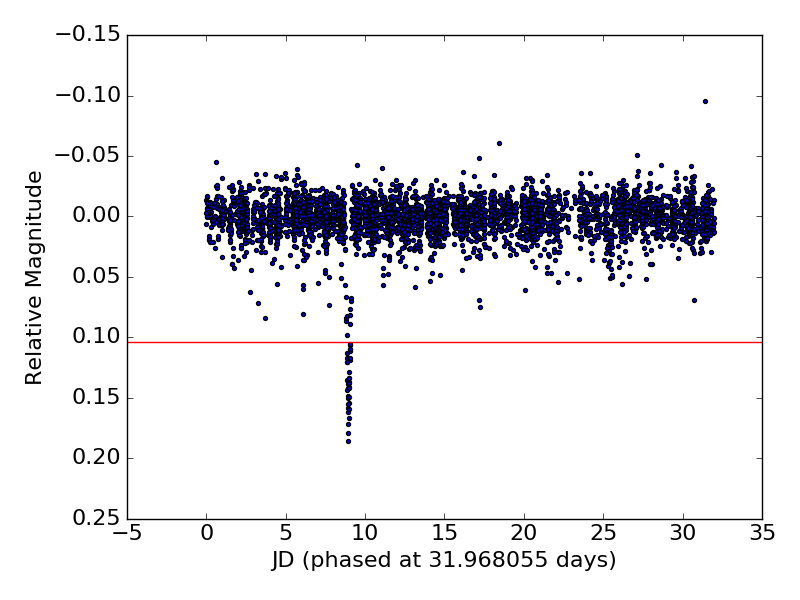}
  \end{center}
  \caption{An eclipsing binary observed by KELT, with the full unphased light curve shown on the left and the phased light curve shown on the right. 15 median absolute deviation outliers are marked in red in the full light curve, and are those below the red line in the phased light curve on right.}
  \label{fig:KELT_demo}
\end{figure*}

We first show both the full light curve and the phased light curve in Figure~\ref{fig:KELT_demo}, with the points in red in the plot of the full light curve being those that we identify as 'in eclipse' by finding all points that are at least 7 standard deviation outliers. In the phased light curve, we mark the outlier threshold with a horizontal red line. In total, there are 24 points that are marked as in-eclipse here. Each possible pairing of 2 points has a unique set of period ranges that can fit those data points, for a total of 552 sets of period ranges for the whole light curve.

As can be roughly seen by eye in Figure~\ref{fig:KELT_demo}, the points that we have identified as in eclipse are spread out over 4 separate consecutive eclipses, and so when we discuss the time difference between any two points in transit, they will be separated by a length in time of between 0 and 3 multiples of the orbital period, modulo up to the duration of the eclipse. This means that our delta time (the time between any two points in eclipse) will be roughly 0, 32, 64, or 96 days. As points that are in the same eclipse do not provide a constraint on period, we focus only on the pairings of points that occur during different epochs.

\begin{figure}[!htb]
  \begin{center}
      \includegraphics[width=0.4\textwidth]{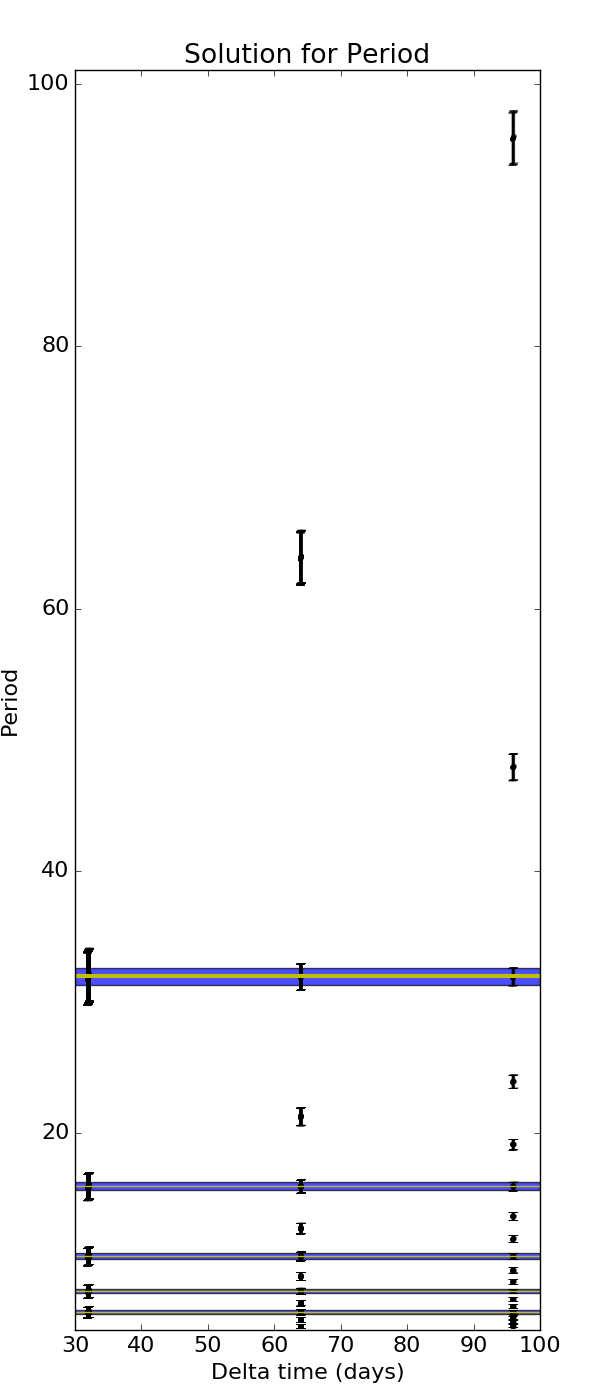}
  \end{center}
  \caption{Each set of error bars at the same x-value represents one set of period ranges that correspond to two points in transit for the KELT EB, with the time between these two points providing the x-value and period ranges being shown on the y-axis. In blue horizontal bars, we mark all regions in period space that are solutions for every pair of points. The yellow horizontal lines represent the harmonics of the true period. All 5 period ranges that are part of our solution coincide with one of these harmonics.}
  \label{fig:KELT_demo2}
\end{figure}

In Figure~\ref{fig:KELT_demo2}, we plot all of the sets of period ranges that we get from the Tingley algorithm for the EB light curve, with a constraint that the duration is no more than 2 days long and the period is no less than 5 days. Pairs of points separated by 3 orbits (96 days) then provide potential solutions centered at around 96 days as well as the harmonic terms for this period (48 days, 32 days, 24 days) down to the minimum period cutoff of 5 days. The period solution for any 2 points separated by about 96 days will have some additional variation as the points occur somewhere within the eclipse, and so are separated by close to, but not necessarily exactly, three times the period. Pairs of points separated by 2 orbits (64 days), similarly will provide solutions close to 64 days, as well as 32 days, 21 days, 16 days, and so on. Finally, the pairs of points separated by just one orbit provide potential solutions of 32 days, as well as near 16 days, 10 days, 8 days, and 6 days. We then look for the periods that are part of the solution sets for all pairs of points.
When we do this, we find 5 period ranges, each of which is between several to tens of hours wide. These are shown as blue horizontal bars in Figure~\ref{fig:KELT_demo2}. For this event to be periodic, the true period must fall in one of these period ranges. In marking the true period and its harmonics with yellow horizontal lines, we see that the top period range, from 31.328 to 32.629 days, brackets the true period of 31.968 days. Additionally, the 4 other period ranges in our solution occur at one half, one third, one fourth, and one fifth the true period and correspond to the harmonics of the true period. In this case, then, the Tingley algorithm constrains the period of this EB to 5 regions of period space, all of which mark the harmonics of the true period.

For this algorithm to be useful in determining a period, we would need at least 3 separate events to be observed. As the wide-fast-deep survey has a minimum revisit time of $\sim$1 hour, much longer than the duration of a planetary transit of a white dwarf, we can treat any light curve with at least three points in transit as suitable for a periodicity search. To determine the outliers in the LSST light curve, we set our cutoff at 15 times the median absolute deviation. The method we use here also becomes computationally intensive as the number of points in transit increases, and so if there are more than ten points in transit, we randomly select ten of them to run the algorithm on. Given that a planet on a transiting orbit is only transiting the planet about 1/1000th of the time, this cutoff is only a factor for white dwarfs that are located in deep-drilling fields where the number of observations is much larger than 1000.
\vfill
\section{Results}\label{results}
\subsection{Example Detection}\label{example}
\begin{figure}[!htb]
  \begin{center}
      \includegraphics[width=0.4\textwidth]{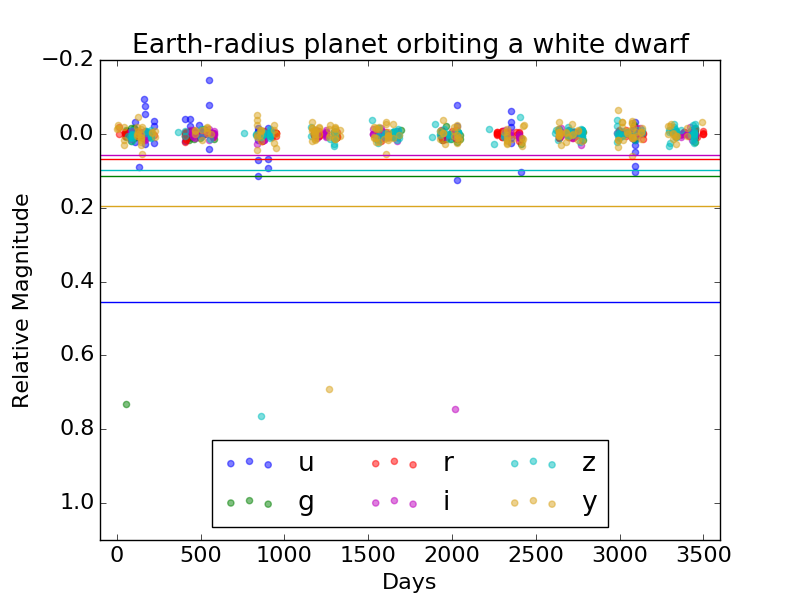}
  \end{center}
  \caption{The full multi-band light curve of a white dwarf with a transiting planet, observed in an LSST wide-fast-deep field. Points are color-coded by band, and the horizontal lines represent the outlier threshold in each band.}
  \label{fig:WD}
\end{figure}

We also provide a more in-depth look at a single transiting system. In Figure~\ref{fig:WD} we show the light curve of one of the white dwarfs with a correctly constrained period, based on observed transits in 4 bands. This white dwarf is observed at a cadence typical of a wide-fast-deep field, with gaps caused by the seasonal observability by LSST clearly visible over the ten-year-long light curve.

\begin{figure*}[!htb]
  \begin{center}
      \includegraphics[width=0.3\textwidth]{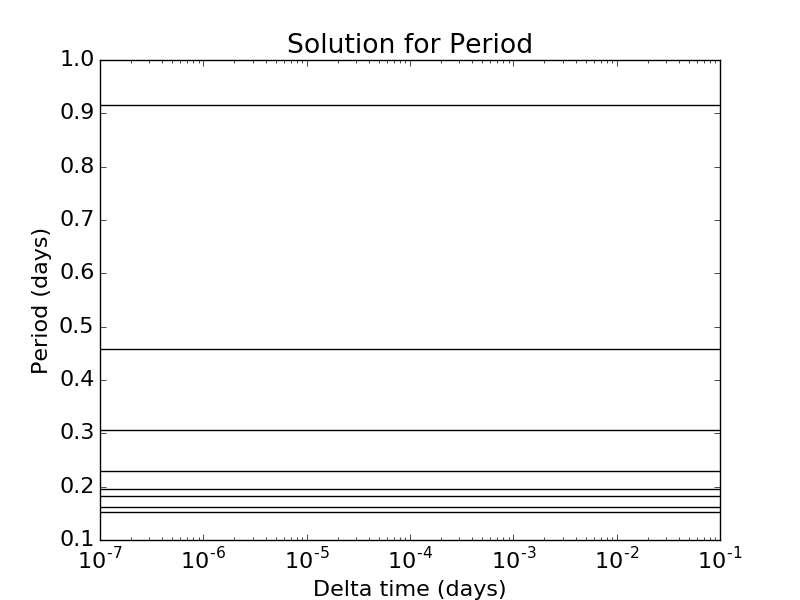}
      \includegraphics[width=0.3\textwidth]{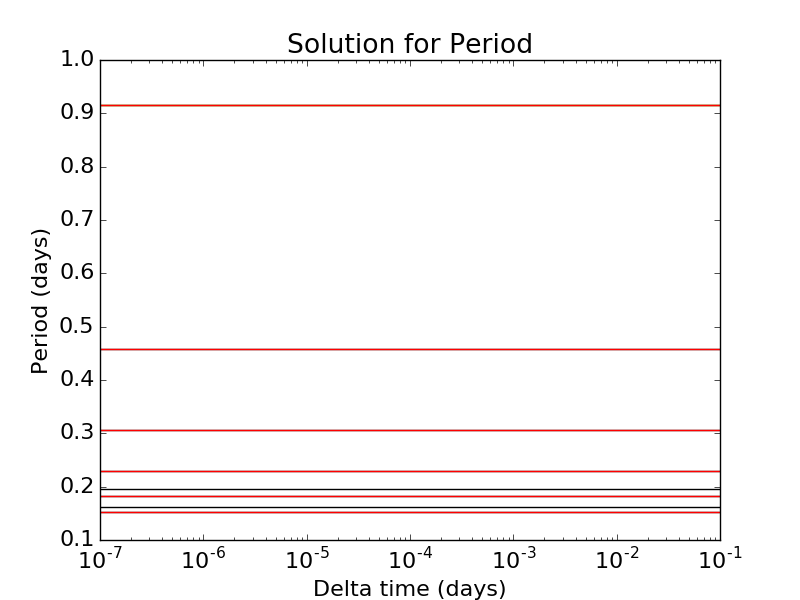}
      \includegraphics[width=0.3\textwidth]{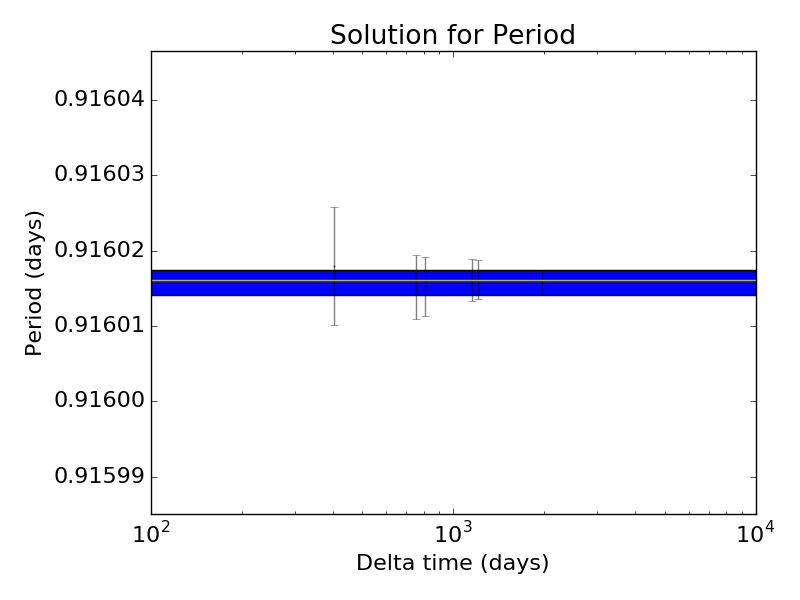}
  \end{center}
  \caption{The family of light curve solutions on the left. In the center, we highlight in red the correct period and all fractional aliases, showing that most of the possible periods fit for this white dwarf are simple aliases of the true period. On the right is a zoomed in view of the input period, showing that there is a tightly constrained region of period space around it. The vertical lines represent each of the intervals between two points in transit.}
  \label{fig:WD_solution}
\end{figure*}

When we look at the solution for the period, presented in Figure~\ref{fig:WD_solution}, we find that there are a very narrow set of possible periods, and that the period has been constrained to being under 0.916 days (this corresponds to the true period). Additionally, when we look at how tightly constrained the period is, the range around the correct period is only 0.3 seconds wide. For a transit taking place one year later, the accuracy for the time of eclipse will still be on the order of 1 minute.

\subsection{Full Results} 
We can look at how many of the 2.7 million white dwarfs we simulate reached each stage of the analysis, by creating light curves for all stars, and injecting transits into those with transiting planets. From the initial population, 11549 white dwarfs had planets on transiting orbits. Because we define transit detection on the outlying data points we can have outliers without an injected transit, and we characterize how common that is.
There were also 17459 white dwarfs that were detected as having at least one outlier data point consistent with a dimming event. 11184 of these were white dwarfs that did not have a transiting planet and just happened to have at least one large outlier. There were 2160 stars that met our stipulation from Section~\ref{sec:ToyModel} of at least 3 outliers observed, 
and all but 22 were transiting systems. When the Tingley algorithm was run on these 2160 systems, we find that in 2132 of the 2138 transiting systems, the correct period is included in the constrained period ranges that are generated as the solution. This represents a recovery rate of 18.5\% for all genuine transiting systems. For 215 of these systems, instead of returning a family of period ranges (as our example in Section~\ref{example} does), we are left with only a single period range, leaving a particularly well-constrained period solution as we do not have to worry about identifying the correct harmonic.
Of the 22 non-transiting systems that had at least 3 points in transit, 18 had solutions returned from the algorithm. Of these, almost all are located in deep-drilling fields. Not only are these systems much more likely to have at least 3 random outliers (as they are observed more than 10 times more frequently than the wide-fast-deep survey), but the current deep-drilling schedule is to observe in 40 one-hour blocks, it is also possible to have 2 points that are separated by less than 5 minutes, and this will invalidate our assumption that if there are three points in transit, they must represent three different transits.

\begin{figure*}[!htb]
  \begin{center}
      \includegraphics[width=0.4\textwidth]{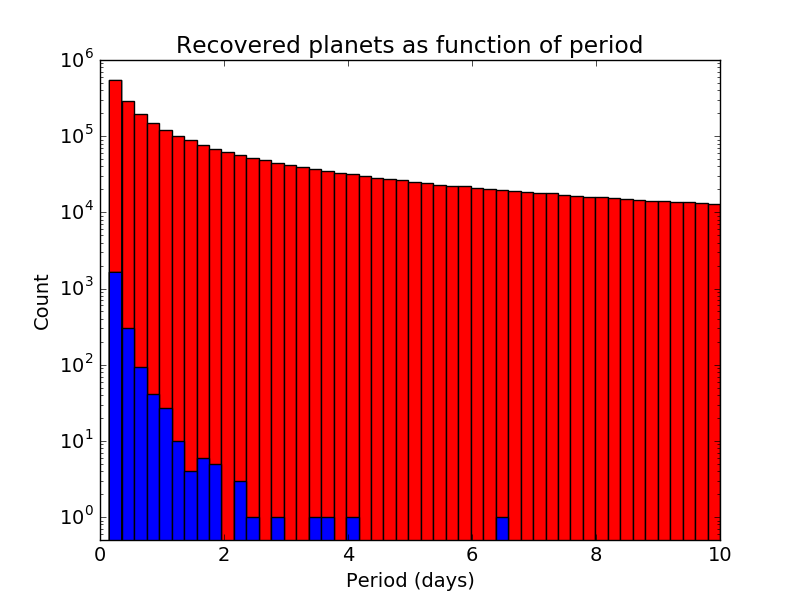}
  \end{center}
  \caption{The distribution of planets in our simulation shown as a function of period. Red represents the total population of simulated planets, and blue represents just those that were recovered. 
  }
  \label{fig:sim_pop}
\end{figure*}

As expected from the work in Section~\ref{sec:approx}, most of the planets that we recover are at shorter periods, as shown in Figure~\ref{fig:sim_pop}, with the entire population of planets shown in red, and the recovered planets shown in blue. In those calculations, the chance of a planet being recovered in any given bin drops below 1 at around 5 days, with the longest period planets recovered in this simulation being a handful of planets that have periods of 2-7 days. The vast majority of planets are recovered at periods below 2 days. We can break down our results by LSST survey, with 2082 planets in the wide-fast-deep survey, 10 planets in the deep-drilling fields, and 40 planets in the remaining subsurveys outside the wide-fast-deep region. While we find almost all planets that had three detected outliers, the number of transiting systems with 3 outliers detected is only about 45\% of the initial estimate in Section~\ref{sec:ToyModel}, significantly lowering the number of planets that we can find. We can also scale our yield to a more conservative planet occurrence rate, and at 0.25 planets per white dwarf, this still represents over 500 planets in the wide-fast-deep survey.

An assumption made in Section~\ref{sec:ToyModel} is that we would have full efficiency in detecting points in transit. However, our simulated population of white dwarfs are all being observed in multiple bands, and faint observations in a given band may not allow for the detection of points in transit at the very high threshold that we have used to avoid incorrectly identifying points in transit. This is a necessary trade-off, as the Tingley algorithm will not work if it includes points that do not occur in transit.

\begin{figure*}[!htb]
  \begin{center}
      \includegraphics[width=0.8\textwidth]{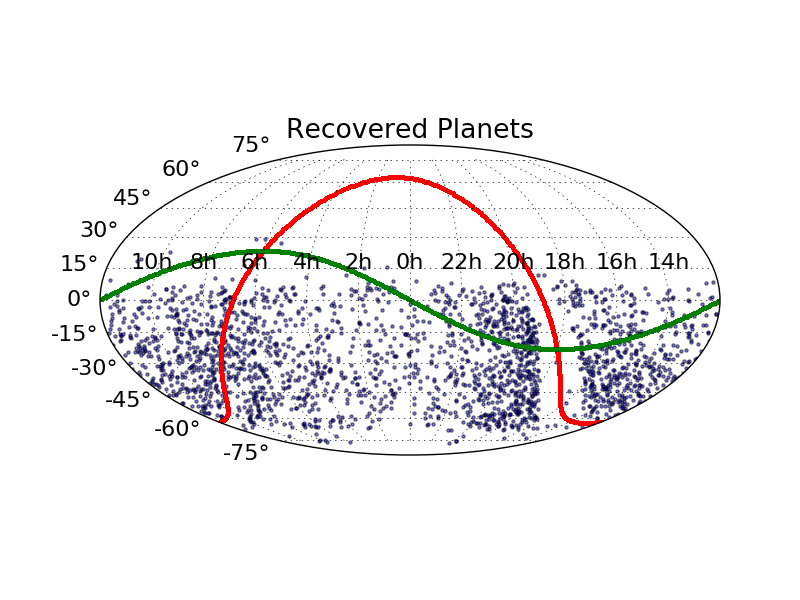}
  \end{center}
  \caption{A map of all recovered planets. At around 18h the very strong impact of the wide-fast-deep survey not including the galactic plane is clearly visible. The galactic plane and the ecliptic are marked in red and green respectively.}
  \label{fig:pop_map}
\end{figure*}

As our initial model presumed 900 observations that are randomly sampled, this only applies to the wide-fast-deep survey. Outside of this region, there will currently only be on order of $\sim$200 observations over ten years, and these regions account for about 40\% of our simulated population of white dwarfs. This includes the galactic plane, where the concentration of white dwarfs is highest, and so the chances of observing white dwarfs in transit will be very negatively impacted by this scheduling decision. In Figure~\ref{fig:pop_map} we show the sky locations of all stars where we recovered a transiting planet. The galactic plane is marked in red, and we can easily see that in the galactic plane, there are almost no planets recovered and the areas with the highest numbers of planets are those just outside of the boundaries for the galactic plane subsurvey (refer to Figure~\ref{fig:LSSTcounts}). We see similar very low rates of detection in the other subsurveys that receive significantly fewer observations, as the northern ecliptic spur and the south celestial pole are also regions with few planets recovered.

\section{Discussion}
A key conclusion from these results is the importance of directed follow-up to confirm the precise periods of these planets. While some planets did have enough observations to determine the period uniquely, most planets will need follow-up observations to determine which harmonic is the true period, as LSST data will not provide a unique solution. Given the short duration of a transit, only 10-20 minutes of follow-up observation time will be needed to confirm or reject a particular period, and so while the constraints on the timing of follow-up observations will be very precise, these follow-up observations will be very quick to carry out.

\begin{figure*}[!htb]
  \begin{center}
      \includegraphics[width=0.4\textwidth]{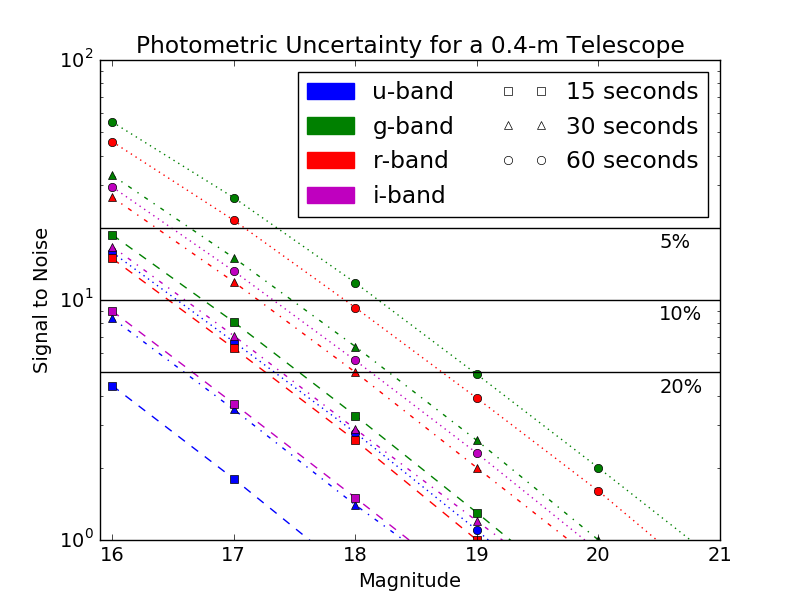}
      \includegraphics[width=0.4\textwidth]{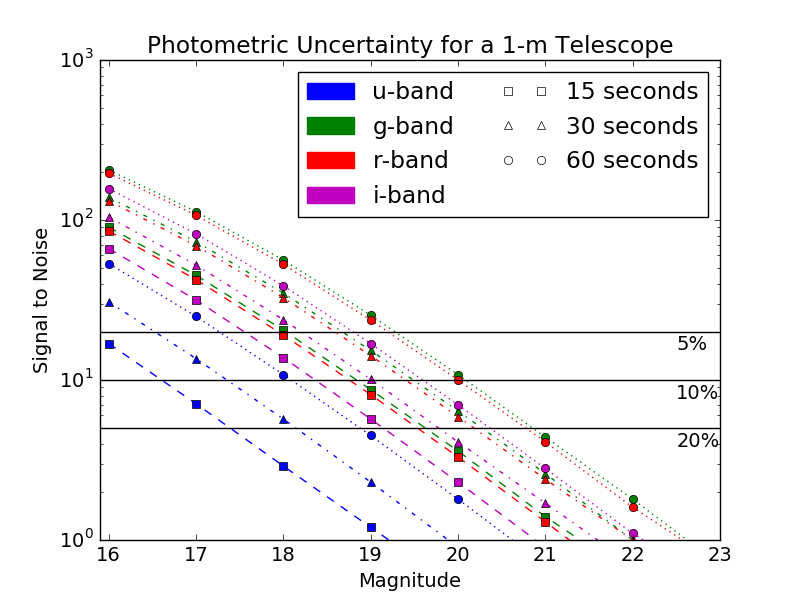}
  \end{center}
  \caption{Signal to noise for white dwarf follow-up observations with 0.4-m (left) and 1-m (right) telescopes, using the Las Cumbres Observatory (LCO) telescope network as a guide. We show the signal-to-noise for the \emph{ugri} bands for 15, 30, and 60 exposures. Bands are represented by color, and exposure length by shape. The horizontal lines mark signal-to-noise of 5\%, 10\%, and 20\%.}
  \label{fig:follow_up}
\end{figure*}

We present a quick demonstration of potential follow-up capability by using the Las Cumbres Observatory (LCO) facilities to provide an example of photometry \citep{Brown2013}. LCO consists of a global network of 0.4-m, 1-m, and 2-m robotic telescopes with automatic scheduling, making it an ideal model for follow-up observations, as only ten to fifteen minutes of observation time would be needed, but with a very significant constraint on the time of observation. Focusing just on the more numerous 0.4 and 1 meter telescopes, we look at the expected signal-to-noise of follow-up observations as a function of magnitude, band, and exposure time using the LCO Exposure Time Calculator \footnote{\url{https://lco.global/files/etc/exposure_time_calculator.html}}. As the time scale for a transit is on order of 5 minutes, we look at exposure times of 15, 30, and 60 seconds. In Figure~\ref{fig:follow_up} we plot the signal-to-noise, with horizontal lines marking the cutoffs for 5\%, 10\%, and 20\% photometric uncertainty, in contrast to a transit depth of $\sim$50\%. The white dwarf with planetary transits discussed in Section~\ref{example} has magnitudes of 22.5 in \emph{u}, 20.7 in \emph{g}, 19.8 in \emph{r}, and 19.4 in \emph{i}. For these magnitudes, a 60-second exposure would be greater than 20\% uncertainty, but a white dwarf about one magnitude brighter would be observable at around 20\% uncertainty with 60-second exposures, and \emph{r} and \emph{g} band observations have 10\% uncertainty for white dwarfs brighter than around 18th magnitude. The \emph{u} band would not provide reliable photometry for our white dwarf at any of our exposure times, with signal-to-noise of less than 1. Our same white dwarf, with the 1-m telescopes, can be observed with better than 10\% uncertainty in the \emph{r} and \emph{i} bands, and about 20\% uncertainty for the \emph{g} band. Better than 5\% uncertainty is possible for white dwarfs down to about 19th magnitude in \emph{gri}, with even 15-second exposures being an option down to about 18th magnitude.

The results here also represent a lower limit in a few ways. For the purposes of this work we have used light curves with a fairly simple model for the transit, with equatorial transits and boxcar transit model. Including grazing transits will potentially increase the number of planets that could be detected, and a more realistic ingress and egress model will increase the likelihood that observations are made in transit by increasing the full duration of a transit event. Both of these factors will need to be understood in order to determine the relationship between the number of detected planets from LSST and what the underlying frequency of planets around white dwarfs is.

The impact of planet size has also not been examined here, and planets larger than 1 Earth radius will cause deeper events and have longer durations, and so while the frequency of larger planets has a much lower upper limit than smaller planets \citep{VanSluijs2017}, it will be important to determine how much easier they would be to detect using this approach. Again, further simulations will be needed in order to determine what constraints can be placed on the underlying exoplanet population distribution as a function of period based on what LSST does, or doesn't detect.

\section{Summary}\label{sec:sum}
In this work, we have explored LSST's ability to detect Earth-sized planets in short-period orbits around white dwarfs, both analytically and with simulations. As planetary transits of a white dwarf can be much deeper than the noise expected in LSST's photometry, we set an initial constraint that any planet that is observed to be transiting at least three times can be detected. We then analytically determine the probability for a planet to transit and to be observed thee times during transit if the star is observed 900 times (equivalent to LSST's wide-fast-deep cadence). When we assume that each white dwarf has a planet drawn from a loguniform distribution between 0.15 and 10 days, we find that of the 1.8 million white dwarfs, about 1 in 400 would have a planet that meets our transit criteria. This amounts to about 4500 planets around white dwarfs.

In our more detailed simulation, we use a more realistic set of obsevational properties for white dwarfs, with a simulated white dwarf population that has been modeled to determine apparent magnitudes and position. 
We combine this information with the LSST OpSim results and a simple boxcar transit model to create simulated light curves for each white dwarf that will accurately represent how LSST will observe them. We implement an algorithm designed for high signal-to-noise transits proposed in \citet{Tingley2011} to then search for periodicity in extreme outliers in the light curve, and demonstrate that this algorithm is very well-suited to detecting transits of white dwarfs. Using our planet rate of 1 close Earth-sized planet per star for our entire population, we find LSST recovering over 2000 planets. At a lower rate of 0.25 planets per white dwarf, this is still over 500 planets. This represents about 20\% of the transiting planets in the sample, or about 1 in every 1200 close-in Earth-sized planets. Of particular note is that it becomes readily apparent that LSST has a very limited ability to discover white dwarfs located in the plane of the galaxy under the current observing strategy of only $\sim$300 observations per star, as the wide-fast-deep survey contains only 60\% of the white dwarfs in our initial population, but 97\% of our recovered planets.

\acknowledgments
The authors thank Stephen Ridgeway, Virginia Trimble, and Jay Farihi for their valuable feedback on this manuscript. The authors would also like to thank J.J. Hermes, Siyi Xu, and Andrew Shannon for productive discussions about white dwarfs, as well as Scott Daniel for assistance in accessing with the LSST CatSim and accessing the University of Washington's CatSim database.

Software: astropy \citep{Price-Whelan2018}, matplotlib \citep{Hunter2007}, numpy \citep{Oliphant2006}, pandas \citep{Mckinney2011}

\bibliographystyle{apalike}
\bibliography{libAAS}

\begin{thebibliography}{}

\bibitem[Agol, 2011]{Agol2011}
Agol, E. (2011).
\newblock {Transit surveys for earths in the habitable zones of white dwarfs}.
\newblock {\em \apj}, 731(2):L31.

\bibitem[Am{\^{o}}res and L{\'{e}}pine, 2005]{Amores2005}
Am{\^{o}}res, E.~B. and L{\'{e}}pine, J. R.~D. (2005).
\newblock {Models for Interstellar Extinction in the Galaxy}.
\newblock {\em \aj}, 130(2):659--673.

\bibitem[Batalha et~al., 2010]{Batalha2010}
Batalha, N.~M., Borucki, W.~J., Koch, D.~G., et~al. (2010).
\newblock {Selection, prioritization, and characteristics of kepler target
  stars}.
\newblock {\em Astrophysical Journal Letters}, 713(2 PART 2).

\bibitem[Bergeron et~al., 1995]{Bergeron1995}
Bergeron, P., Wesemael, F., and Beauchamp, A. (1995).
\newblock {Photometric Calibration of Hydrogen- and Helium-Rich White Dwarf
  Models}.
\newblock {\em \pasp}, 107:1047.

\bibitem[Brown et~al., 2013]{Brown2013}
Brown, T.~M., Baliber, N., Bianco, F.~B., et~al. (2013).
\newblock {Las Cumbres Observatory Global Telescope Network}.
\newblock pages 1031--1055.

\bibitem[Catanzarite and Shao, 2011]{Catanzarite2011}
Catanzarite, J. and Shao, M. (2011).
\newblock {The occurrence rate of earth analog planets orbiting Sun-like
  stars}.
\newblock {\em \apj}, 738(2).

\bibitem[Connolly et~al., 2014]{Connolly2014}
Connolly, A.~J., Angeli, G.~Z., Chandrasekharan, S., et~al. (2014).
\newblock {An end-to-end simulation framework for the Large Synoptic Survey
  Telescope}.
\newblock 9150:915014.

\bibitem[Connolly et~al., 2010]{Connolly2010}
Connolly, A.~J., Peterson, J., Jernigan, J.~G., et~al. (2010).
\newblock {Simulating the LSST system}.
\newblock Number August 2010, page 77381O.

\bibitem[Debes and Sigurdsson, 2002]{Debes2002}
Debes, J.~H. and Sigurdsson, S. (2002).
\newblock {Are There Unstable Planetary Systems around White Dwarfs?}
\newblock {\em \apj}, 572(1):556--565.

\bibitem[Dressing and Charbonneau, 2015]{Dressing2015}
Dressing, C.~D. and Charbonneau, D. (2015).
\newblock {the Occurrence of Potentially Habitable Planets Orbiting M Dwarfs
  Estimated From the Full Kepler Dataset and an Empirical Measurement of the
  Detection Sensitivity}.
\newblock {\em \apj}, 807(1):45.

\bibitem[Faedi et~al., 2011]{Faedi2011}
Faedi, F., West, R.~G., Burleigh, M.~R., et~al. (2011).
\newblock {Detection limits for close eclipsing and transiting substellar and
  planetary companions to white dwarfs in the WASP survey}.
\newblock {\em \mnras}, 410(2):899--911.

\bibitem[Gaudi et~al., 2017]{Gaudi2017}
Gaudi, B.~S., Stassun, K.~G., Collins, K.~A., et~al. (2017).
\newblock {A giant planet undergoing extreme-ultraviolet irradiation by its hot
  massive-star host}.
\newblock {\em \nat}, pages 1--39.

\bibitem[Harrison et~al., 2018]{Harrison2018}
Harrison, J. H.~D., Bonsor, A., and Madhusudhan, N. (2018).
\newblock {Polluted white dwarfs: constraints on the origin and geology of
  exoplanetary material}.
\newblock {\em \mnras}, 479(3):3814--3841.

\bibitem[Hunter, 2007]{Hunter2007}
Hunter, J.~D. (2007).
\newblock {Matplotlib: A 2D graphics environment}.
\newblock {\em Computing in Science and Engineering}, 9(3):99--104.

\bibitem[Ivezic et~al., 2008]{Ivezic2008}
Ivezic, Z., Tyson, J.~A., Acosta, E., et~al. (2008).
\newblock {LSST: from Science Drivers to Reference Design and Anticipated Data
  Products}.
\newblock page~34.

\bibitem[Jacklin et~al., 2015]{Jacklin2015}
Jacklin, S.~R., Lund, M.~B., Pepper, J., et~al. (2015).
\newblock {Transiting Planets with LSST II. Period Detection of Planets
  Orbiting 1 Solar Mass Hosts}.
\newblock {\em \aj}, 150(1):34.

\bibitem[Jacklin et~al., 2017]{Jacklin2017}
Jacklin, S.~R., Lund, M.~B., Pepper, J., et~al. (2017).
\newblock {Transiting Planets with LSST. III. Detection Rate per Year of
  Operation}.
\newblock {\em \aj}, 153(4):186.

\bibitem[Jones et~al., 2014]{Jones2014a}
Jones, R.~L., Yoachim, P., Chandrasekharan, S., et~al. (2014).
\newblock {The LSST metrics analysis framework (MAF)}.
\newblock volume 9080, page 91490B.

\bibitem[Juri{\'{c}} et~al., 2008]{Juric2008}
Juri{\'{c}}, M., Ivezi{\'{c}}, {\v{Z}}., Brooks, A., et~al. (2008).
\newblock {The Milky Way Tomography with SDSS. I. Stellar Number Density
  Distribution}.
\newblock {\em \apj}, 673(2):864--914.

\bibitem[Kilic et~al., 2013]{Kilic2013a}
Kilic, M., Agol, E., Loeb, A., et~al. (2013).
\newblock {Habitable Planets Around White Dwarfs: an Alternate Mission for the
  Kepler Spacecraft}.
\newblock pages 1--11.

\bibitem[Livio et~al., 2005]{Livio2005}
Livio, M., Pringle, J.~E., and Wood, K. (2005).
\newblock {Disks and Planets around Massive White Dwarfs}.
\newblock {\em \apj}, 632(1):L37--L39.

\bibitem[Loeb and Maoz, 2013]{Loeb2013a}
Loeb, A. and Maoz, D. (2013).
\newblock {Detecting biomarkers in habitable-zone earths transiting white
  dwarfs}.
\newblock {\em \mnras : Letters}, 432(1):L11--L15.

\bibitem[{LSST Science Collaboration} et~al.,
  2017]{LSSTScienceCollaboration2017}
{LSST Science Collaboration}, Marshall, P., Anguita, T., et~al. (2017).
\newblock {Science-Driven Optimization of the LSST Observing Strategy}.
\newblock pages 1--312.

\bibitem[Lund et~al., 2015]{Lund2014}
Lund, M.~B., Pepper, J., and Stassun, K.~G. (2015).
\newblock {Transiting Planets With Lsst. I. Potential for Lsst Exoplanet
  Detection}.
\newblock {\em \aj}, 149(1):16.

\bibitem[Mazeh et~al., 2012]{Mazeh2012}
Mazeh, T., Nachmani, G., Sokol, G., et~al. (2012).
\newblock {Kepler KOI-13.01 – Detection of beaming and ellipsoidal
  modulations pointing to a massive hot Jupiter}.
\newblock {\em \aap}, 541:A56.

\bibitem[Mckinney, 2011]{Mckinney2011}
Mckinney, W. (2011).
\newblock {pandas : a Foundational Python Library for Data Analysis and
  Statistics}.
\newblock (January 2011).

\bibitem[Nutzman and Charbonneau, 2008]{Nutzman2008}
Nutzman, P. and Charbonneau, D. (2008).
\newblock {Design Considerations for a Ground-Based Transit Search for
  Habitable Planets Orbiting M Dwarfs}.
\newblock {\em \pasp}, 120(865):317--327.

\bibitem[Oliphant, 2006]{Oliphant2006}
Oliphant, T.~E. (2006).
\newblock {\em {Guide to NumPy}}.
\newblock Trelgol Publishing.

\bibitem[Pepper et~al., 2007]{Pepper2007}
Pepper, J., Pogge, R.~W., DePoy, D.~L., et~al. (2007).
\newblock {The Kilodegree Extremely Little Telescope (KELT): A Small Robotic
  Telescope for Large‐Area Synoptic Surveys}.
\newblock {\em \pasp}, 119(858):923--935.

\bibitem[Price-Whelan et~al., 2018]{Price-Whelan2018}
Price-Whelan, A.~M., Sipőcz, B.~M., G{\"{u}}nther, H.~M., et~al. (2018).
\newblock {The Astropy Project: Building an Open-science Project and Status of
  the v2.0 Core Package}.
\newblock {\em \aj}, 156(3):123.

\bibitem[Shporer et~al., 2011]{Shporer2011}
Shporer, A., Jenkins, J.~M., Rowe, J.~F., et~al. (2011).
\newblock {Detection of KOI-13.01 using the photometric orbit}.
\newblock {\em \aj}, 142(6).

\bibitem[Shporer et~al., 2014]{Shporer2014}
Shporer, A., O'Rourke, J.~G., Knutson, H.~A., et~al. (2014).
\newblock {Atmospheric characterization of the hot Jupiter Kepler-13Ab}.
\newblock {\em \apj}, 788(1).

\bibitem[Silvotti et~al., 2014]{Silvotti2014a}
Silvotti, R., Sozzetti, A., Lattanzi, M., et~al. (2014).
\newblock {Detectability of substellar companions around white dwarfs with
  Gaia}.
\newblock pages 1--4.

\bibitem[Tingley, 2011]{Tingley2011}
Tingley, B. (2011).
\newblock {Searching for transits in data with long time baselines and poor
  sampling}.
\newblock {\em \aap}, 529:A6.

\bibitem[van Lieshout et~al., 2018]{VanLieshout2018}
van Lieshout, R., Kral, Q., Charnoz, S., et~al. (2018).
\newblock {Exoplanet recycling in massive white-dwarf debris discs}.
\newblock 28(May):1--28.

\bibitem[van Sluijs and {Van Eylen}, 2018]{VanSluijs2017}
van Sluijs, L. and {Van Eylen}, V. (2018).
\newblock {The occurrence of planets and other substellar bodies around white
  dwarfs using K2}.
\newblock {\em \mnras}, 474(4):4603--4611.

\bibitem[Vanderburg et~al., 2015]{Vanderburg2015}
Vanderburg, A., Johnson, J.~A., Rappaport, S., et~al. (2015).
\newblock {A disintegrating minor planet transiting a white dwarf}.
\newblock {\em \nat}, 526(7574):546--549.

\bibitem[Veras et~al., 2013]{Veras2013}
Veras, D., Mustill, A.~J., Bonsor, A., et~al. (2013).
\newblock {Simulations of two-planet systems through all phases of stellar
  evolution: implications for the instability boundary and white dwarf
  pollution}.
\newblock {\em \mnras}, 431(2):1686--1708.

\bibitem[Villaver and Livio, 2007]{Villaver2007}
Villaver, E. and Livio, M. (2007).
\newblock {Can Planets Survive Stellar Evolution?}
\newblock {\em \apj}, 661(2):1192--1201.

\bibitem[Wolszczan and Frail, 1992]{Wolszczan1992}
Wolszczan, a. and Frail, D.~a. (1992).
\newblock {A planetary system around the millisecond pulsar PSR1257 + 12}.
\newblock {\em \nat}, 355(6356):145--147.

\bibitem[Xu and Jura, 2012]{Xu2012}
Xu, S. and Jura, M. (2012).
\newblock {Spitzer observations of white dwarfs: The missing planetary debris
  around DZ stars}.
\newblock {\em \apj}, 745(1).

\end{thebibliography}

\end{document}